\documentclass{PoS}

\title{Distortion of the HBT image by mean field interaction}

\ShortTitle{Distortion of the HBT image by mean field interaction}

\author{\speaker{Koichi Hattori}
\\
\\
Institute of Physics and Applied Physics, Yonsei University, Seoul 120-749, Korea
\\
\\
Institute of Particle and Nuclear Studies, High Energy Accelerator Research Organization (KEK), 1-1 Oho, Tsukuba, Ibaraki 305-0801, Japan
\footnote{Former affiliation at the time of the conference}
\\
E-mail: \email{khattori@post.kek.jp}}

\abstract{
We study effects of a mean-field interaction on the spacetime geometry of the hadron source 
measured by utilizing the Hanbury Brown and Twiss (HBT) interferometry in the ultrarelativistic heavy-ion collisions. 
We show how a modification of a pion amplitude, caused in the freeze-out process, is incorporated into the correlation function 
of the interferometry within a semiclassical method. 
Profiles of the distorted images are illustrated. 
To make a quantitative estimate of the effects, we construct a mean-field-interaction model on the basis of the pion-pion scattering amplitude, 
and then investigate to what extent the effects of the mean-field interaction acts to efficiently modify the HBT radii. 
}

\FullConference{The Seventh Workshop on Particle Correlations and Femtoscopy\\
		 September 20th to 24th, 2011\\
		 The University of Tokyo, Japan}

\newcommand{\bx}{{\bm{x}}}

\newcommand{\bk}{{\bm{k}}}
\newcommand{\bp}{{\bm{p}}}
\newcommand{\bq}{{\bm{q}}}

\newcommand{\integ}{\int\!\!}

\newcommand{\dS}{\delta \! S^{c\ell}_{\bk}}

\newcommand{\out}{_{\rm out}}
\newcommand{\side}{_{\rm side}}

\usepackage{bm}
\usepackage{graphicx}
\usepackage{amsmath,amssymb,mathrsfs}
\usepackage{cases}

\begin{document}

\section{Introduction}

\begin{figure}[t]
 \begin{minipage}{0.52\hsize}
  \begin{center}  
  \includegraphics[scale=0.35]{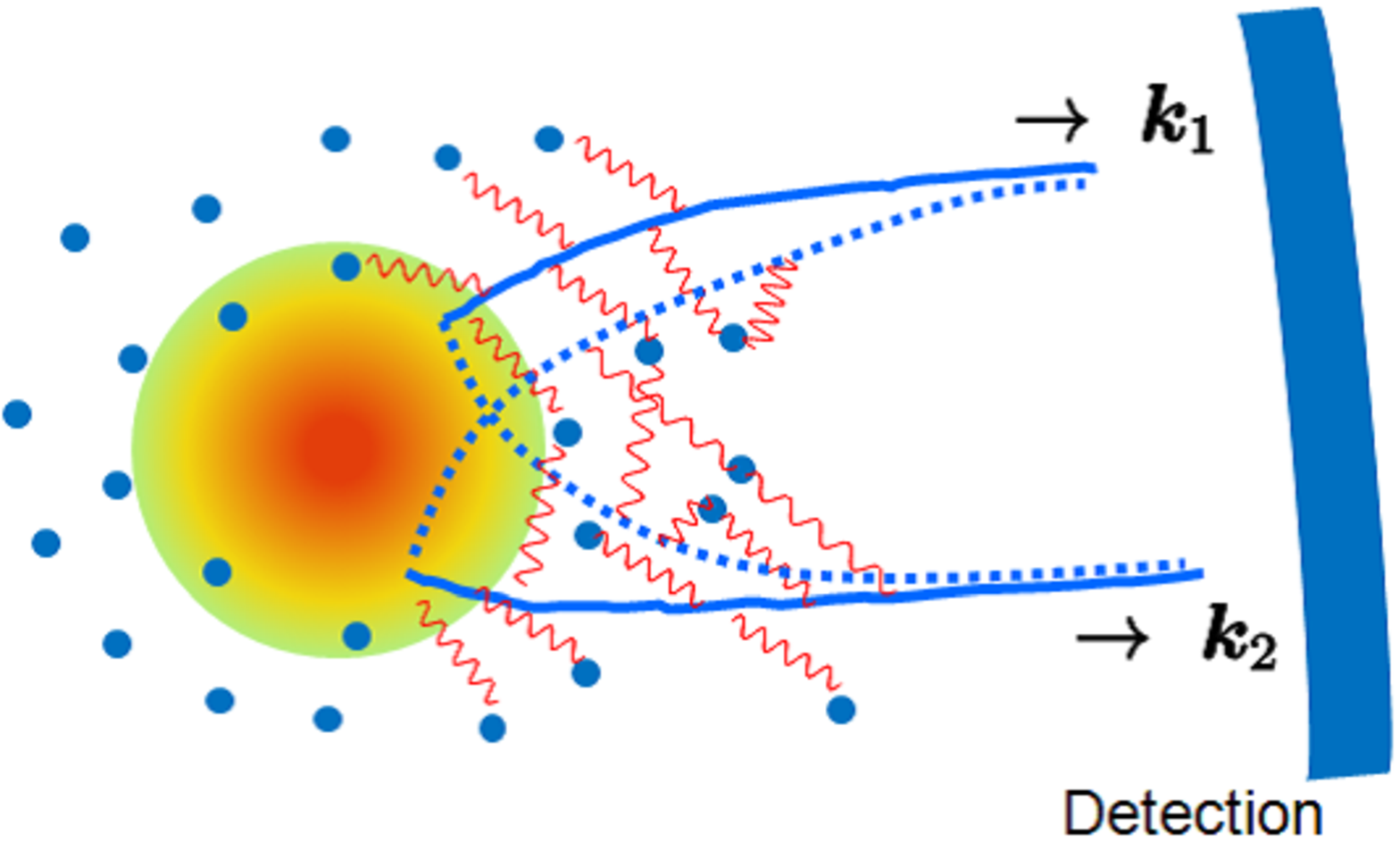}  
  \end{center}
\vspace{-0.5cm}
\caption{An emitted particle interacts with a mean-field formed by other evaporating particles. 
This figure also shows an interference between two possible trajectories of an identical particle pair. }
\label{fig:MF}
 \end{minipage}
 \hspace{0.1cm}
 \begin{minipage}{0.455\hsize}
  \begin{center}  
  \includegraphics[scale=0.25]{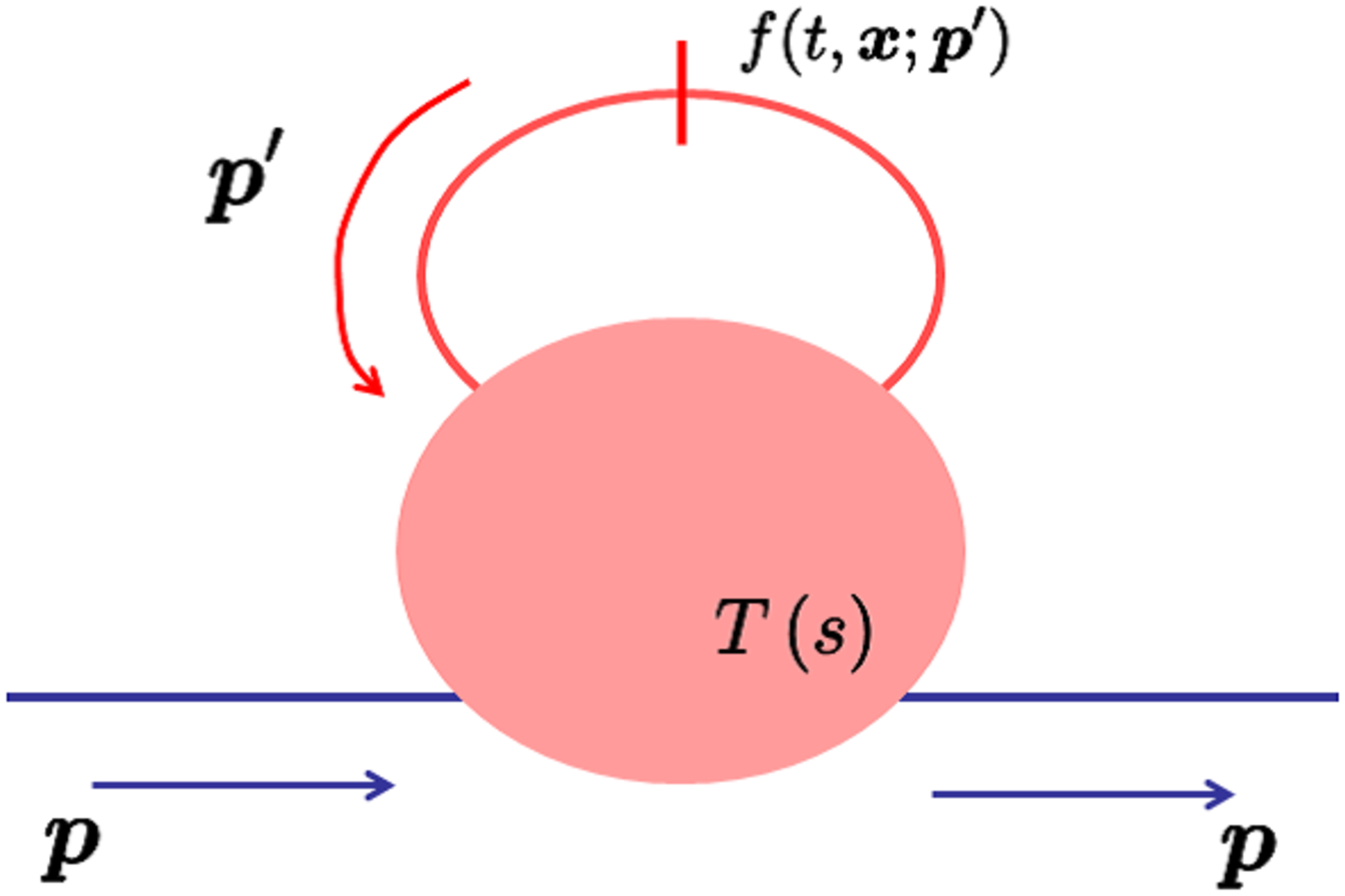} 
  \end{center}
\caption{Phenomenological pion self-energy: a propagating pion is scattered by other evaporating pions represented with a loop. 
Blob indicates the forward scattering amplitude shown in Fig. \protect \ref{fig:pipi}.}
\label{fig:se}
 \end{minipage}
 \hspace{0.2cm}
\end{figure}

In prior to the RHIC experiments, the dynamics of the collective expansion was simulated with hydrodynamical models assuming a formation of the quark-gluon plasma. 
Their results indicated a prolonged lifetime of the matter, 
and a considerable difference in the transverse HBT radii, $R\out/R\side \simeq 2$, which has been supported by sophisticated models\cite{LPSW}. 
While these models have precisely reproduced the single-particle spectra, especially the elliptic flow, of the RHIC data, 
it was unexpectedly found that there were systematic discrepancies between the theoretically predicted and experimentally measured HBT radii. 
It has been known as the ``RHIC HBT puzzle". 
This contribution is based on the works \cite{HM,KH} originally motivated by solving this puzzle. 
We investigate effects of a mean-field interaction, sketched in Fig. \ref{fig:MF}, as a possible origin of the discrepancies. 
\footnote{
Effects of mean-field interactions have been examined independently in Ref. \cite{CMWY} and \cite{Pra06} to the end of solving the puzzle.  
We thank H. Fujii for calling our attention to those works. 
}.

We have shown how the modification of a pion amplitude due to a mean-field interaction 
reflects in a distorted HBT image\cite{HM}, within a semiclassical framework. 
We also found a possible improvement of the discrepancies, 
owing to cooperative effects of attractive and absorptive mean-field interactions.  
To pursue this possibility, we have constructed a more realistic mean-field interaction 
on the basis of the pion-pion forward scattering amplitudes\cite{KH}. 
There, we actually obtained attractive and absorptive mean-field interactions 
following from a strong attraction and attenuation around the momentum regime of the $\rho$-meson resonance. 
Based on those frameworks, we showed how and to what extent the mean-field interaction efficiently modify the HBT radii.

\section{Modification of the correlation function}


We show how the effects of the mean-field interaction, encoded in the phase and attenuation factors of the amplitude, 
reflect in the interference pattern. 
By using a semiclassical method to evaluate the amplitude, 
a modified correlation function has been obtained in a time-independent case\cite{HM}, 
and then the framework has been extended to a time-dependent case to take into account the expansion dynamics of the hadron source\cite{KH}.

The correlation function $C$ in the HBT interferometry is defined by 
\begin{eqnarray}
C(\bm{k}_1,\bm{k}_2)  = \frac{ P_2(\bm{k}_1,\bm{k}_2) }{ P_1(\bm{k_1}) P_1(\bm{k_2}) } \ ,
\label{eq:C} 
\end{eqnarray}
where $\bm{k}_1$ and $\bm{k}_2$ are the momenta of a particle pair. 
$P_1(\bm{k})$ and $P_2(\bm{k}_1,\bm{k}_2)$ are probabilities of detecting a particle and identical particle pair, respectively. 
Without any interaction after the emission, the correlation function is simply given by the Fourier transform of the source function $S(x, \bk) $ as, 
\begin{eqnarray}
C(\bk, \bq) = 1 + \eta^2(\bk )
\left|  \integ d^4x \ S(x, \bk) \ e^{i q x } \right|^2 \ .
\label{eq:C0}
\end{eqnarray}
The correlation function is measured as a function of the average and relative momenta, $\bk$ and $\bq$. 
The product of the single-particle spectra appears as a normalization factor, 
$
\eta^{-2}(\bk) \simeq P_1^2(\bk) = \left[ \ \integ d^4x \ S(x, \bk) \ \right] ^2
\ . 
$
In this expression, we have approximated the momenta as $\bm k_1 \simeq \bm k_2 \simeq \bm k$ for the small relative momentum $( |\bq| \ll |\bk|)$, 
which is a relevant regime for the HBT interferometry.

Once we incorporate effects of the mean-field interaction, they distort the kernel of the Fourier transform in Eq. (\ref{eq:C0}). 
In terms of the classical action of a pion propagating in the mean field, 
a modified correlation function is obtained as, 
\begin{eqnarray}
C (\bk, \bq) = 1 + \eta^{2} (\bk) 
\left| 
\integ d^4x  \ S(x, \bp_0 (x,\bk) ) \ 
e^{-2\gamma(x,\bk)} \ 
e^{ i q_{\mu} \left( x^{\mu} + \partial_k ^{\mu} \delta \! S^{c\ell}_{\bk} (x) \right) } 
\right| ^2 
\ , \label{eq:C1}
\end{eqnarray}
where $x^{\mu}=(t,\bx)$ is the emission point, and the derivative $\partial_k^\mu$ operates on the four momentum $k^\mu$. 
The mean-field interaction acts to accelerate the pion emitted with an initial momentum $\bp_0 (x,\bk)$ so that it becomes the asymptotic one $\bk$. 
An attenuation factor is given by the coefficient $\gamma(x,^bk)$ 
which is related to an imaginary part of pion self-energy (see Sec.\ref{sec:MF})
\footnote{
We extended the semiclassical framework to take into account the relativity in an appendix in Ref. \cite{KH}. 
}. 
The deviation of the classical action from that of the free propagation, $\dS = S^{c\ell}_{\bk} - k^\mu x_\mu $, appears as the phase distortion. 
Reflecting the acceleration and attenuation in the mean field, 
the single-particle spectrum is also modified as, 
$\eta^{-2} (\bk) = \left[ \integ d^4x  \ S(x, \bp_0 (x,\bk) ) \ e^{-2\gamma(x,\bk)} \right] ^2 $.

\begin{figure}[t]
  \begin{center}  
  \includegraphics[scale=0.57]{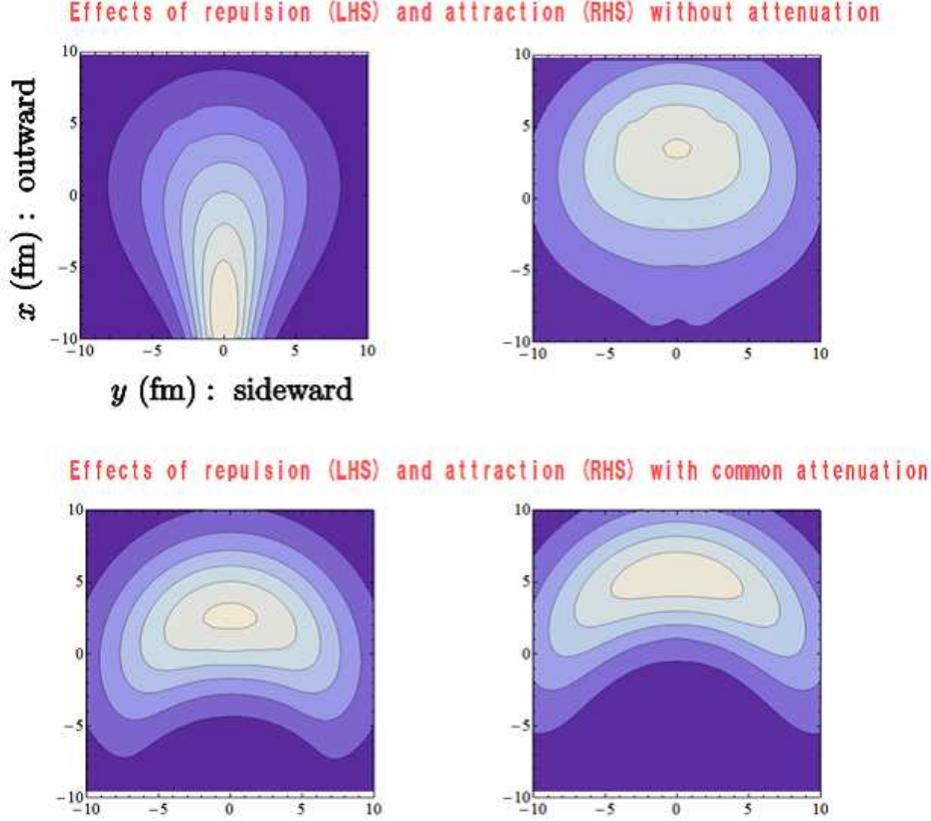}  
  \end{center}
\vspace{-.2cm}
\caption{Distorted HBT images at $\bk_\perp = 100$ MeV: 
contour plots show how an isotropic Gaussian profile, with its variance $\sigma = 5 \ {\rm fm}$, is apparently distorted due to effects of repulsive and attractive interactions, 
and cooperative effects of attenuation, as indicated above. 
}
\label{fig:profile}
\end{figure}

In Fig. \ref{fig:profile}, we illustrate the distorted HBT images 
due to the mean-field interactions (see Ref.\cite{HM} for details). 
A repulsive mean-field interaction induces an image elongated along the outward axis, 
whereas an attractive one induces a stretched image with a longer extension along the sideward axis. 
Effects of an attenuation also acts to stretch the sideward extension, 
because it cuts off a propagation of the pion emitted on the opposite side to the detection point.

\section{Phenomenological mean-field interaction} \label{sec:MF}

In the isospin-symmetric limit, an in-medium pion self-energy depicted in Fig. \ref{fig:se} is written as, 
\begin{numcases}
{}
\Pi(t, \bx, \bp) = -  \integ \frac{d^4p^\prime}{(2\pi)^4} \   
2\pi \delta(p^{\prime 2}-m^2) \ T(s) \  f(t, \bx,\bp^\prime)   \label{eq:SE} \\
p^{\prime\mu}\partial_\mu f(t, \bx,\bp^\prime) = S(t, \bx,\bp^\prime) 
\label{eq:Bol}
\end{numcases}
with the isospin averaged scattering amplitude $T(s)$ and distribution of the medium pion $f(t, \bx,\bp^\prime)$. 
A Mandelstam variable $s = (p+q)^2$ is the squared center-of-mass energy of a two-body scattering of on-shell pions. 
Eq. (\ref{eq:Bol}) describes the time-evolution of the distribution function in the presence of a source term $S(t, \bx,\bp^\prime)$ 
which take into account the pion emission from the hadron source. 
We take the source function being common to the one in Eq. (\ref{eq:C1}) as, 
\begin{eqnarray}
S(t, \bx, \bp) = \frac{1}{(2\pi)^3} 
\int f_{eq}(x^{\prime}, p)\  \delta^{_(4)}(x-x^{\prime}) \ p^{ \mu} d\sigma_{\mu}(x^\prime) \ , \label{eq:source}  
\end{eqnarray}
where $x^\prime$ is the emission point on the hypersurface, of which normal vector is given by $d\sigma_{\mu}(x^\prime)$. 
The thermal distribution function $f_{eq}(x, p) = \left[\exp \{ (p^{ \nu} u_{\nu}-\mu)/T \} -1 \right]^{-1}$ is specified with 
the macro variables: temperature $T$, pion chemical potential $\mu$ 
and the flow vector $u^{\mu} \propto \left( \frac{t}{\tau}, \bm{v}_{\perp}, \frac{z}{\tau} \right)$ with the normalization $u^\mu u_\mu = 1$. 
As a simple model, we take $ T=130 \ {\rm MeV}$ and $\mu = 30 \ {\rm MeV}$, 
and assume Bjorken flow and the transverse flow profile proportional to the transverse coordinate $r$ as $\bm v_\perp = 0.06 r \ {\it c}^{-1}$ 
(see Ref \cite{KH} for details of the model).

\begin{figure}[t]
  \begin{center}  
  \includegraphics[scale=0.5]{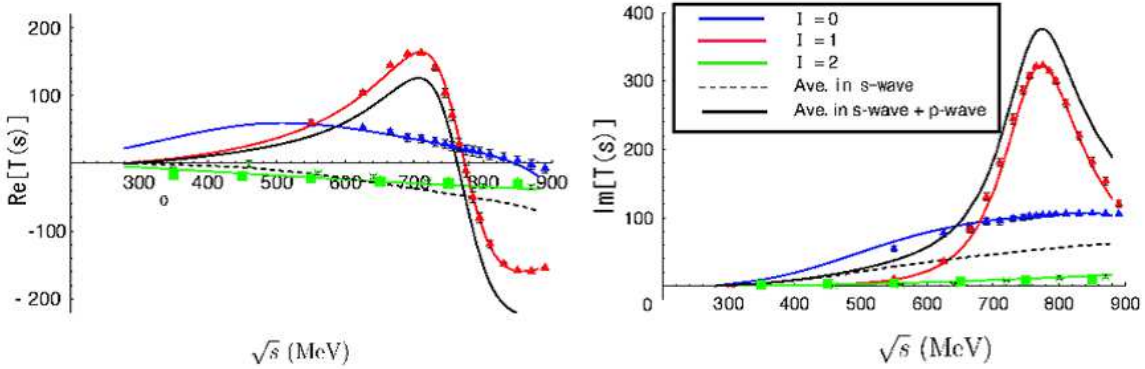}  
  \end{center}
\vspace{-.8cm}
\caption{Real and imaginary parts of the forward pion-pion scattering amplitude\cite{PI}. 
Effects of the $\rho$-meson resonance manifest in the mean-field interaction as an attraction and attenuation.}
\label{fig:pipi}
\end{figure}

The forward pion-pion scattering amplitude (\ref{eq:SE}) is given by averaging over the isospin channels as 
$T(s) = 3 \left( \frac{1}{9} T_0 (s) + \frac{3}{9}T_1 (s) + \frac{5}{9}T_2 (s) \right) $. 
The overall factor comes from the contributions of the pion spices, $\pi^\pm$ and $\pi^0$, 
assuming the same emission rate (\ref{eq:source}) for each. 
Panels in Fig. \ref{fig:pipi} show the real and imaginary parts of the isospin-dependent scattering amplitudes and their average, respectively. 
A strong attraction and attenuation around the $\rho$-meson resonance efficiently modify a pion amplitude. 
Because effects of the mean-field interaction (\ref{eq:SE}) depend on a path and momentum of a pion propagation, 
pions acquire phase shifts and attenuations in different magnitudes, depending on their motions after the emissions. 
Therefore, they cannot maintain the interference compared with the case without the mean-field interaction. 


\section{Results}

\begin{figure}[t]
  \begin{center}  
  \includegraphics[scale=0.49]{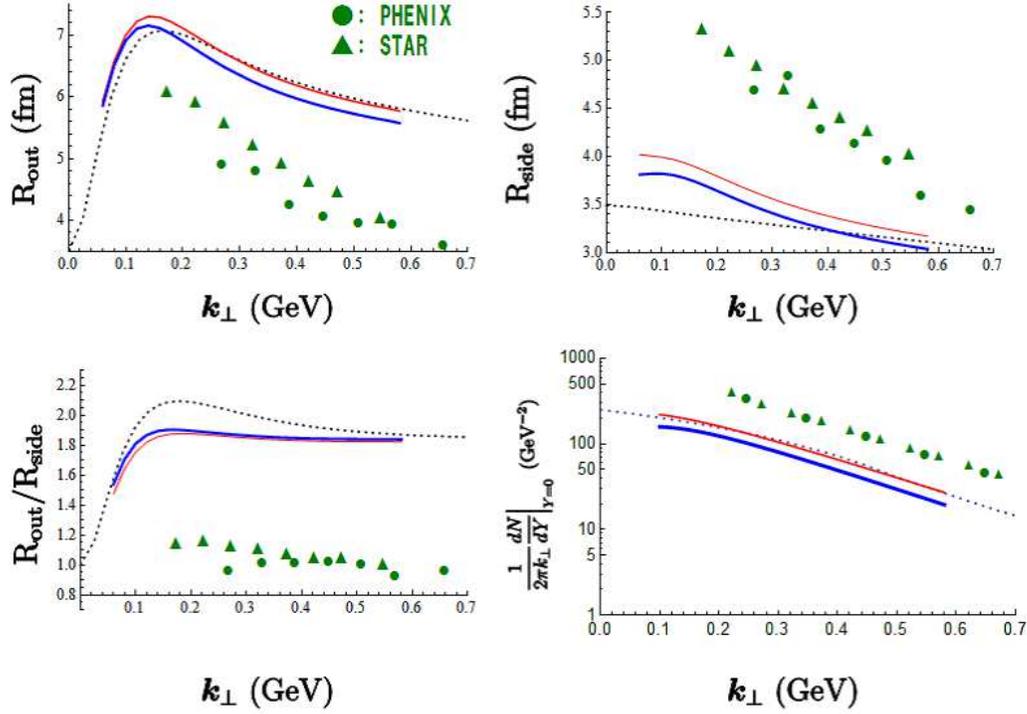}  
  \end{center}
  \vspace{-.5cm}
\caption{Transverse HBT radii and single-pion spectrum: 
red curves show these quantities in the presence of the real part of the self-energy, 
while blue curves show cooperative effects of the real and imaginary parts of the self-energy. 
Dotted curves show the quantities obtained without the mean-field interaction. 
}
\label{fig:result}
\end{figure}

In Fig. \ref{fig:result}, we show effects of the mean-field interaction on the transverse HBT radii and single-pion spectrum. 
Those effects appear as displacements from dotted curves (see caption) which imitate typical results of the hydrodynamical simulations. 
We find that the mean-field interaction mostly modify the sideward radius 
and that it acts to reduce a discrepancy between the dotted curve and the experimental data 
which is seen in typical results of conventional hydrodynamical simulations. 
Owing to this improvement, a deviation in the ratio of the transverse radii is also improved in a low momentum regime. 

The effect of the real part do not cause a considerable shift of the single-pion spectrum, 
and an amount of attenuated pions noticed by the blue curve would be totally complemented 
if we consistently take into account a decay of the $\rho$-meson resonance as an inverse reaction of the attenuation. 
This is a good point for a consistent description of the transverse HBT radii and single-pion spectrum 
in a sense of improving the discrepancies found in the hydrodynamic picture: 
otherwise precisely reproduced single-pion spectrum would be sacrificed.

\section{Concluding remarks}

We would like to remark on a prospect toward a consistent description of the hadron phase. 
It would be worthwhile to investigate the effects of the mean-field interaction throughout the hadron phase by more elaborate descriptions, 
because the interaction between an emitted pion and a rather dense hadron source would be more effective, 
and because the magnitudes of the effects obtained in our work are 
comparable to those of the individual effects obtained with several upgrades of hydrodynamical modeling\cite{Pra09}. 
An analytic study of a mean-field interaction in terms of kinetic theory will be found in Ref. \cite{MM}.

\section*{Acknowledgments}

The author thanks the organizers and especially S. Pratt as the convener of the session 
for giving him an opportunity to talk in the conference. 
A large part of this contribution is based on his Ph.D. thesis submitted in 2010. 
He is grateful to his supervisor, Prof. T. Matsui, for his encouraging instruction.


\end{document}